\documentclass[aps,pre,twocolumn,showpacs]{revtex4}

\usepackage[dvips]{graphicx}
\usepackage{psfrag}
\usepackage{amsmath}
\usepackage{amssymb}

\begin{document}

\title{Comment on ``Existence of Internal Modes of Sine-Gordon Kinks''}
\author{C. R. Willis}
\affiliation{Department of Physics, Boston University, 590
  Commonwealth Avenue, Boston Massachusetts 02215}

\begin{abstract} 
In Ref.[1] [Phys. Rev. B. {\bf 42}, 2290 (1990)] we used a rigorous
projection operator collective variable  formalism for nonlinear Klein-Gordon
equations to prove the continuum  Sine-Gordon (SG) equation has a long lived
quasimode whose frequency $\omega_s$= 1.004 $\Gamma_0$ is  in the continuum
just above the lower phonon band edge with a lifetime ($1/\tau_s$ ) = 0.0017
$\Gamma_0$. We confirmed the analytic calculations by simulations which
agreed  very closely with the analytic results. In Ref.[3]
[Phys. Rev. E. {\bf 62}, R60 (2000)] the authors performed  two numerical
investigations which they asserted ``show that neither intrinsic internal
modes nor quasimodes exist in contrast to previous results.'' In this paper
we prove their first numerical investigation could not possibly observe the
quasimode in principle and their second numerical investigation actually
demonstrates the existence of the SG quasimode. Our analytic calculations
and verifying simulations were performed for a stationary Sine-Gordon
soliton fixed at the origin. Yet the authors in Ref.[3] state the
explanation of our analytic simulations and confirming simulations are due
to the Doppler shift of the phonons emitted by our stationary Sine-Gordon
soliton which thus has a zero Doppler shift.
\end{abstract}
\pacs{05.45.Yv, 03.50.-z}
\maketitle

\section{INTRODUCTION}

In Ref.\cite{Boesch90} we proved the continuum Sine-Gordon (SG) equation has
a long  lived quasimode whose frequency from simulation is $\omega_s = (1.004
\pm 0.001)\,\Gamma_0$ (where $\Gamma_0$ is  the frequency of the lower band
edge in units where the speed of sound $c$ = 1) and whose lifetime from
simulation is $(1/\tau_s ) = (0.003 \pm 0.001)\,\Gamma_0$ . We used a  rigorous
projection operator collective variable (CV) formalism for  nonlinear
Klein-Gordon equations derived in Ref.\cite{Boesch88} to calculate the
quasimode frequency and lifetime. Our calculated theoretical values for the 
frequency and inverse lifetime are $\Omega = 1.00585\,\Gamma_0$ and 
$(1/\tau_s) = 0.0017\,\Gamma_0$ which agree very well with our simulation values.

In Ref.\cite{Quintero00} the authors performed two numerical investigations
for the SG in  which they assert ``show that neither intrinsic internal modes
nor  quasimodes exist in contrast to previous results'' referring to
Ref.\cite{Boesch90}. In Secs.III and IV we analyze their two numerical
investigations in detail and prove their first numerical investigation could
not possible observe the quasimode in principle and that their second
numerical investigations actually observes the SG quasimode at the beginning
of their simulation. However, the length of their system was so short for
their long observation time, that there were many transversals of the system
by phonons emitted at different times by the soliton, which then reflected
from the end of the system and then interfered with phonons emitted
later. Thus each phonon interfered with phonons emitted earlier and phonons
emitted later which led to a very complicated interference pattern. The
authors of Ref.\cite{Quintero00} concluded that the complicated interference
pattern was a ``proof'' that SG quasimodes don't exist. However, the correct
conclusion is that their poorly designed numerical investigation was for a
time t that was more than ten times too long for the length of their system
to avoid the irrelevant interferences. In the first 200 seconds just before
the first emitted phonons reflected off the end of the system and returned
to the stationary emitting SG soliton, the finite lifetime of the quasimode
is clearly  observable.

Our analytic calculations and verifying simulations were all for a
continuum, force free and stationary SG soliton i.e., the center of mass of
the SG was fixed at the origin for all times. However, in their two
numerical investigations in Ref.\cite{Quintero00} the author provided an
explanation of  our analytic calculations and verifying simulations which was
that our  phonons were Doppler shifted. Which is truly amazing for phonons
emitted by  a stationary SG soliton fixed at the origin.
As a result, their two numerical investigations and their ``explanation'' of 
our results have absolutely no relevance to the validity of our analytic 
soliton and verifying simulations of our continuum stationary and force free 
SG quasimode.

In Sec.II we outline the derivation of the exact equations of motion for
the SG equation. We prove in Sec.III that the first numerical search for
the SG quasimode in Ref.\cite{Quintero00} could not observe the SG quasimode
in principle. While in Sec.IV we show that in their second numerical
investigation the authors of Ref.\cite{Quintero00} actually observe the SG
quasimode at the beginning of their simulation. However, their simulation
was for a time, too large for the length of their system. Consequently, they
observed a complicated interference pattern which was totally irrelevant in
the SG quasimode mode that was clearly observable at the beginning of their
simulation. In Sec.VI we present our conclusions and discuss a recent work,
Ref.\cite{Kalbermann04}, which contains a new solution of the SG equation by
using the inverse transform method and find our SG quasimode solution is
valid.

\section{EQUATIONS OF MOTION FOR THE SG QUASIMODE}

The purpose of this section is to outline the derivation of the equations of
motion from our exact CV equations for the SG equation in
Ref.\cite{Boesch90} that we actually solved for the SG quasimode in
Ref.\cite{Boesch90}. We need these rigorous equations in order to contrast
them with the two CV equations of motion for the kink momentum $P(t)$ and
width of the kink $l(t)$ which form the basis of their theoretical
analysis of our equations of motion and which we repeat below in Eqs.(6)
and (8). Where we show their equation of motion for $P(t)$ is totally
irrelevant to our derivations of the SG quasimode and their equation of
motion for $\Gamma(t)$ is independent of the phonon dressing whose
interaction with $\chi$ gives rise to the SG quasimode. The slope $\Gamma(t)
= 2 \pi \,[\,l(t)\,]^{-1}$.

We start with the equation of motion for $X(t)$, $\Gamma(t)$, and $\chi(t)$ whose solutions 
are rigorously equivalent to the solution of the SG partial differential 
equation
\begin{equation}
\frac{\partial^2\phi}{\partial t^2} - \frac{\partial^2\phi}{\partial x^2} +
\bigg[\,\frac{\pi}{l_0}\,\bigg]^2 \,\sin\phi = 0,
\end{equation}
where 
\begin{equation}
\phi(x,t) = \sigma[\,\xi(t)\,] + \chi[\,\xi(t),t\,],
\end{equation}
and 
\begin{equation}
\xi(t) \equiv \Gamma(t)\,[\,x-X(t)\,].
\end{equation}
The soliton solution $\sigma$ is
\begin{equation}
\sigma(\xi) = 4\,\tan^{-1}\exp[\xi].
\end{equation}
$\chi(t)$ is the solution of
\begin{align}
\frac{\partial^2\chi}{\partial t^2} -\,& \chi''\bigg[\,\Gamma^2(1-\dot{X}^2) +
2\,\xi\,\dot{X}\,\dot{\Gamma} -
\bigg[\frac{\dot{\Gamma}}{\Gamma}\,\bigg]^2\,\xi^2\,\bigg]\notag \\
+\,& 2\,\frac{\partial\chi'}{\partial
  t}\bigg[\,\bigg[\frac{\dot{\Gamma}}{\Gamma}\,\bigg]\,\xi -
\dot{X}\,\Gamma\,\bigg]\notag \\
+\, & \chi'\,\bigg[\,\bigg[\frac{\ddot{\Gamma}}{\Gamma}\,\bigg]\,\xi
-2\,\dot{X}\,\dot{\Gamma} - \ddot{X}\,\Gamma\,\bigg]\, + \frac{\partial
  V(\sigma+\chi)}{\partial\sigma}\notag \\
=\, & \sigma''\bigg[\,\Gamma^2(1-\dot{X}^2) + 2\,\xi\,\dot{X}\,\dot{\Gamma}
-\bigg[\frac{\dot{\Gamma}}{\Gamma}\bigg]^2\,\xi^2\,\bigg]\notag \\
-\, & \sigma'\,\bigg[\,\bigg[\frac{\ddot{\Gamma}}{\Gamma}\,\bigg]\,\xi
-2\,\dot{X}\,\dot{\Gamma} - \ddot{X}\,\Gamma\,\bigg].
\end{align}  
The equation of motion for $\ddot{X}(t)$ is
\begin{align}
\ddot{X}=
-\,&\frac{\dot{X}\,\dot{\Gamma}}{\Gamma(1-b_X)}-\frac{1}{M_X(1-b_X)}\,\bigg[\,
\langle\sigma'|\chi''\rangle \Gamma^2(1-\dot{X}^2)\notag \\
+\,& 2\langle\sigma'|\xi\,\chi''\rangle\dot{X}\,\dot{\Gamma} 
-\bigg[\,\frac{\dot{\Gamma}}{\Gamma}\,\bigg]^2
\langle\sigma'|\xi^2\,\chi''\rangle -2\frac{\dot{\Gamma}}{\Gamma}\,\bigg\langle
\sigma'|\xi\,\frac{\partial\chi'}{\partial t}\bigg\rangle \notag \\
+\,& 2\,\dot{X}\,\Gamma\bigg\langle \sigma'\bigg|\frac{\partial\chi'}{\partial t}
\bigg\rangle -\frac{\ddot{\Gamma}}{\Gamma}\langle\sigma'|\xi\,\chi'\rangle
+2 \langle\sigma'|\chi'\rangle \dot{X}\,\dot{\Gamma} \notag \\ 
-\,&\bigg\langle\sigma'\bigg|\frac{\partial V}{\partial\sigma}\bigg\rangle\,\bigg],   
\end{align}
where 
\begin{equation}
M_X \equiv \Gamma\,\langle\sigma'|\sigma'\rangle
\end{equation}
is the bare mass of the kink associated with the $X$ motion, and
$b_X\equiv\Gamma\,\langle\sigma''|\chi\rangle/M_X$. The dot product $\langle
f|g\rangle$ is defined as
$$
 \langle f|g\rangle \equiv \int\,f^*(\xi)\,g(\xi)\,d\xi.
$$
The equation of motion for $\Gamma(t)$ is
\begin{align}
\ddot{\Gamma}=\,&\frac{3\,\dot{\Gamma}^2}{2\,\Gamma(1-b_\Gamma)}-
\frac{M_X(1-\dot{X}^2)}{2\,\Gamma\,M_\Gamma(1-b_\Gamma)}\notag \\
+\,&\frac{1}{M_\Gamma(1-b_\Gamma)}\,\bigg[\,\langle\xi\,\sigma'|\chi''\rangle
(1-\dot{X}^2)+\frac{2\,\dot{X}\,\dot{\Gamma}}{\Gamma^2}\langle\xi\,\sigma'|\chi''\,
\xi\rangle\notag \\
-\,&\frac{1}{\Gamma^2}\bigg[\,\frac{\dot{\Gamma}}{\Gamma}\,\bigg]^2\langle\xi\,\sigma'|\xi^2\chi''
\rangle-\frac{2}{\Gamma^2}\frac{\dot{\Gamma}}{\Gamma}\bigg\langle\xi\,\sigma'\bigg|\xi\,
\frac{\partial\chi'}{\partial t}\bigg\rangle \notag \\
+\,&\frac{2\,\dot{X}}{\Gamma}\bigg\langle\xi\,\sigma'\bigg|\frac{\partial\chi'}{\partial
  t}\bigg\rangle + \bigg[\,\frac{2\,\dot{X}\,\dot{\Gamma}}{\Gamma^2} +
\frac{\ddot{X}}{\Gamma}\,\bigg]\langle\xi\,\sigma'|\chi'\rangle\notag \\
-\,&\frac{1}{\Gamma^2}\bigg\langle\xi\,\sigma'\bigg|\frac{\partial
  V}{\partial\sigma}\bigg\rangle,      
\end{align}
where
\begin{equation}
M_\Gamma \equiv \Gamma^{-3}\, \langle\xi\,\sigma'|\xi\,\sigma'\rangle.
\end{equation}

Since the center of mass motion does not play any role in the existence of
the SG quasimode we set $X(t)\equiv 0$ in the equations of motion for
$\chi(t)$ and $\Gamma(t)$.  We are interested in small oscillations of the
quasimode so we linearize  Eq.(5) and Eq.(8) to first order in $\chi$ and
obtain
\begin{align}
\frac{\partial^2\chi}{\partial t^2} - &\,\chi''\bigg[\,\Gamma^2 -
\bigg[\,\frac{\dot{\Gamma}}{\Gamma}\,\bigg]^2\xi^2\,\bigg]
+2\frac{\dot{\Gamma}}{\Gamma}\,\frac{\partial\chi'}{\partial
  t}\,\xi+\frac{\ddot{\Gamma}}{\Gamma}\,\xi\,\chi' \notag \\
+&\,\Gamma_0^2\,\sin\sigma + \Gamma_0^2\,\chi\,\cos\sigma =
\sigma''\bigg[\,\Gamma^2 -
\bigg[\,\frac{\dot{\Gamma}}{\Gamma}\,\bigg]^2\,\xi^2 \,\bigg] \notag \\
-&\,\xi\,\sigma'\,\frac{\ddot{\Gamma}}{\Gamma},
\end{align}
and
\begin{align}
\ddot{\Gamma} =&\,\bigg[\,\frac{3\,\dot{\Gamma}^2}{2\,\Gamma} -
\frac{M_X}{2\,\Gamma\,M_\Gamma}\, \bigg](1+b_\Gamma) +
\frac{\langle\xi\,\sigma'|\chi''\rangle}{M_\Gamma}\notag \\
-&\,\frac{1}{M_\Gamma\,\Gamma^2}
\bigg[\,\frac{\dot{\Gamma}}{\Gamma}\,\bigg]^2\,\langle\xi\,\sigma'|\xi^2\chi''\rangle
-\frac{2}{M_\Gamma\,\Gamma^2}\frac{\dot{\Gamma}}{\Gamma}\,\bigg\langle\xi\,\sigma'\bigg|\xi
\frac{\partial\chi'}{\partial t}\bigg\rangle \notag \\
-&\,\frac{\Gamma_0^2}{M_\Gamma\,\Gamma^2}\,\langle\xi\,\sigma'|\sin\sigma\rangle(1+b_\Gamma)
-\frac{\Gamma_0^2}{M_\Gamma\,\Gamma^2}\,\langle\xi\,\sigma'|\chi\,\cos\sigma\rangle. 
\end{align}
Since we are considering only small oscillations in $\chi$ we further linearize 
Eqs.(10) and (11) in $\delta\Gamma \equiv \Gamma(t)-\Gamma_0$. 
Finally we obtain
\begin{equation}
\frac{\partial^2\chi}{\partial t^2} - \Gamma_0^2\,\chi'' +
\Gamma_0^2\,\chi\cos\sigma_0  =
2\,\Gamma_0\,\delta\Gamma\,\sigma_0''-\xi_0\,\sigma_0'\,\frac{\delta\ddot{\Gamma}}{\Gamma_0},
\end{equation}
and
\begin{equation}
\delta\ddot{\Gamma}= -\Omega_{\rm SG}^2\delta\Gamma +
\frac{1}{M_{\Gamma_0}}\langle\xi_0\,\sigma'_0|\chi''\rangle
-\frac{1}{M_{\Gamma_0}}\langle\xi_0\,\sigma'_0|\chi\cos\sigma_0\rangle,  
\end{equation}
where 
$$
\sigma_0 \equiv \sigma\big|_{\Gamma=\Gamma_0}, \;\;\;\;\; \xi_0\equiv \Gamma_0\,x.
$$
In the remainder of Ref.\cite{Boesch90} we solved these equations of motion 
analytically and calculated the lifetime of the quasimode.

\section{FIRST NUMERICAL INVESTIGATION}
The first numerical search for the SG quasimode in Ref.\cite{Quintero00}
consisted of  trying to find the SG quasimode by measuring numerically the
absorption  spectrum of a discrete SG equation driven by an ac field. What
they measured  was the phonon absorption spectrum of the linearized discrete
SG equation in  an ac field which is:
\begin{align}
\ddot{\chi}_n - (\chi_{n+1} +&\,\chi_{n-1} -2\,\chi_{n}) +
\Gamma_0^2\,\chi_n\notag \\ 
=&\,(\sigma_{n+1} +\sigma_{n-1}-2\,\sigma_{n})\dot{X}^2 + f(t),
\end{align}  
where $\sigma_n$ is the discrete SG soliton at position $n$, and $\chi_n$ is
the discrete SG phonon  at position $n$. The external ac field is $f(t) =
\epsilon\,\exp(i\omega t)$ and $\dot{X}(t)$ is the  velocity of the center of
mass of the SG soliton. We point out again our  derivations and simulations
were for a stationary continuum SG where $\dot{X}(t) \equiv  0$. The spectrum
they obtained by numerically solving the ac driver discrete  SG equation is
given in their Fig.(1) namely
\begin{equation}
\omega_n = \bigg[\,1 +
\bigg(\frac{2\,\pi\,n}{l}\bigg)^2\,\bigg]^{1/2}\;\;\;\;\; {\rm for}\;\;
n=1,2,\ldots,N\equiv/\Delta X,
\end{equation} 
which is just the spectrum one obtains by solving Eq.(1) analytically. For
$f(t)=\epsilon\cos\omega t$ and $\dot{X}(t)= 0$ the spectrum is
$\sum_n\,\delta(\omega -\omega_n)$. If you include $X(t)$ you get exactly the
same spectrum by taking $f(t) =\cos[(\omega/2)\,t]$ because $P^2(t)$ is then
proportional to $\cos[2(\omega/2)t]$, which also yields the identical
spectrum $\sum_n\,\delta(\omega -\omega_n)$ which is what they observe and
what one obtains by analytically solving Eq.(14). Strangely Eq.(14) which is
the basis of their first numerical investigation, is never mentioned in
Ref.\cite{Quintero00} only the numerically observed spectrum Eq.(15) is
presented.

What is most important about their first numerical investigation is that it
could not possibly detect the quasimode even in principle. In order to
observe a quasimode in absorption it must first be created and then observed
during its finite lifetime. Consequently, a quasimode is usually observed in
emission. The SG quasimode can be excited as an initial condition by
deforming the slope or the width of the kink as an initial condition as we
did in our derivations and simulations in Ref.\cite{Boesch90} and as the
authors of  Ref.\cite{Quintero00} did in their second numerical investigation
which we discuss in the  next section. The quasimode can also be excited by
any potential that  distorts the slope of width of the SG soliton. The force
on the slope $\Gamma(t)$  due to a potential $V(x)$ is
\begin{equation}
F = \int\, dx \, V(x)\,\frac{\partial\sigma}{\partial\Gamma}.
\end{equation}
For an ac field $V= f(t)$, so the force $F$ vanishes because $f(t)\int\, dx
\,(\partial\sigma/\partial\Gamma) = 0$. Thus an  ac field cannot possibly
excite a phonon mode. Consequently, their first  numerical investigation in
Ref.\cite{Quintero00} could not possibly detect the presence  of the SG
quasimode and thus it has no relevance whatsoever to the existence  or
nonexistence of the SG quasimode. It is important to stress that a  quasimode
is different than an eigenmode of a linearized Klein-Gordon  equation in that
an unoccupied eigenmode exists even if it is unoccupied.  Whereas a quasimode
has first to be created in order to be observed and it  lasts only for its
lifetime.

\section{SECOND NUMERICAL INVESTIGATION}
In Ref.\cite{Boesch90} we performed simulations that verified our analytic
solutions  for the SG quasimode. We performed three simulations for
$\Gamma(t)$ and $\chi(t)$ for a  stationary SG soliton. We also simulated the
Fourier transform of $\Gamma(t)$ which  gives the quasimode frequency and
lifetime. The simulations agree very closely with analytic results. We
considered cases where the initial slope was different than $\Gamma_0$ i.e.,
$\delta\Gamma(0)\neq 0$ and for $\delta\dot{\Gamma}(0)\neq 0$. We considered
two cases where the length of the system was 1000 units and a third case
where the length of the system was 200 units. For the system of length 1000
we followed the time development of $\Gamma(t)$ and $\chi(t)$ for times $t$
that were short compared with the time a spontaneously emitted phonon would
travel to the end of the system reflect and interfere with phonons emitted
later. In the third case we took a short system $L$ = 200 and followed the
time until the spontaneously emitted phonon first reflected from the end of
the system. We pointed out that eventually the first emitted phonons would
reflect form the end and interfere with phonons emitted later. Thus
simulation of the stationary SG quasimode should have a sufficiently long
system that there are no reflections during the time of observation or
equivalently for a fixed length $L$ the time of observation should be less
than $(L/c)$ where $c$ is the speed of sound.

In Fig.(2) of Ref.\cite{Quintero00} the length of the system was $L$ =
100. They followed the time development of the width $l(t)$ for 2500
seconds. The round trip time of a phonon emitted by a stationary SG soliton
at one end reflect and go back to the stationary SG soliton, is 200
seconds. Thus during their 2500 second observation time there were phonons
that made more than twelve trips that would interfere with phonons emitted
earlier and later. The quasimode lifetime is 500 seconds. So phonons could be
emitted, travel to the end of the system, reflect and be reabsorbed by the
still excited quasimode. Consequently, during their 2500 second simulation
time, phonons are continuously being emitted, interfering with previously
emitted phonons reflecting from the ends of the system and sometimes being
absorbed by the stationary SG soliton at the end of the system. Consequently,
the simulation of Fig.(2) in Ref.\cite{Quintero00} should show a very
complicated interference pattern with multiple time scales but with a period
of 200 seconds playing a prominent role, which is precisely what they
observe. If they had taken a much longer system or had just simulated for
times up to 200 seconds, they would have verified the existence of the SG
quasimode. Actually, the first 200 seconds of their simulation of the width
$l(t)$ gives a very good representation of the SG quasimode. Thus the
incompetent design of their simulation in Fig.(2) is the cause of their
meaningless, complicated interference pattern.

Once again, in their second numerical solution the authors of
Ref.\cite{Quintero00} completely ignore the phonon dressing which gives rise
to the SG quasimode. In Ref.\cite{Boesch90} we calculated and simulated the
phonon dressing. Which shows how the dressing decays as the quasimode emits
phonons during its lifetime while the slope decays from $\Gamma(0)$ to the
constant slope $\Gamma_0$ and $\dot{\Gamma}(0)$ decays to zero. Also, as in
their first numerical simulation, the second numerical solution is for an
appreciably discrete phonon system while our derivations and simulations
were for the continuum SG equation. Here its interesting to observe that the
qualitative behavior of the appreciably discrete SG system is similar to our
analytic calculations and simulations for the continuum SG.

\section{DISCUSSION}
The authors of Ref.\cite{Quintero00} performed two numerical investigations
for which  they state ``we show that neither intrinsic internal modes nor
``quasimodes'' exist in contrast to previous reports'' referring in
particular to our Ref.\cite{Boesch90}. In Sec.III we proved that the SG
quasimodes that we had derived analytically and verified by simulation could
not possibly be observed by their first numerical investigation. The reason
is that in order to observe the SG quasimode it must first be created and
then observed during its finite lifetime. We proved in Sec.III that an ac
driver can not create a SG quasimode and thus their ac absorption numerical
experiment could not possibly observe the SG quasimode but could only measure
the phonon absorption spectra of their discrete SG phonon eigenmodes. Our
derivations were, for a force free, stationary, continuum SG soliton.
However, their first numerical investigation is for an ac driven discrete SG
soliton. They measured numerically the discrete SG spectrum. However, the
phonon spectrum of the SG plays absolutely no role in our analysis, doesn't
appear in any of our derivations and is totally irrelevant to our
results. The SG quasimode comes from the solution of the coupled continuum
equations for the slope of the kink, $\Gamma(t)$, and the phonon dressing for
$\chi(t)$. Consequently, their first numerical investigation has no relevance
whatsoever to the existence or nonexistence of the SG quasimode.

In their second numerical investigation of the SG equation they started a
discrete stationary SG with an initial rate of change of the slope
$\dot{\Gamma}(0)\neq 0$ and $\Gamma(0)\neq 0$. In  Ref.\cite{Boesch90} we
considered three such cases except our derivations and simulations were for
the continuum SG and for initial values $\delta\dot{\Gamma}(0)$ of 0.01 and
0.001 and $\delta\Gamma_0 =$ 0.1. The quasimode we derived  analytically and
verified by simulation was for a linear mode. However, they actually observed
the quasimode in the first 200 seconds of their simulation. They however took
a system too short for the length of time they followed the
simulations. Consequently, they obtained a very complicated phonon
interference pattern due to the multiple phonon interferences due to the
earlier emitted phonons interfering with phonons emitted earlier and later
because of the multiple reflections of the phonons from the ends of the
system. In addition, there were multiple absorptions and reemissions of the
phonons with the stationary soliton. If they had increased their system from
$L$ = 100 to $L$ = 400 and followed the simulation for $t$ = 500 seconds
instead of their $t$ = 2500 seconds they would have obtained essentially the 
same diagram we obtained for $\Gamma(t)$.

The authors of Ref.\cite{Quintero00} state their theoretical analysis of our paper is 
based on their two cc equations:
\begin{equation}
\frac{dP}{dt} = -q\,\epsilon\,\sin(\delta t + \delta_0),
\end{equation}
where
$$
P(t) \equiv M_0\,l_0\,\dot{X}\,l^{-1}(t),
$$
and
\begin{equation}
\alpha\,(\,\dot{l}^2 - 2\,l\,\ddot{l}\,) =\frac{l^2}{l_0^2}\,(1+\frac{P^2}{M_0^2})
-1.
\end{equation}
Their width variable $l(t)/l_0$ is essentially the inverse of our variable
$\Gamma(t)$. Since our $X(t)\equiv 0$ their variable $X(t)$ should be
identically zero and have no relevance to any of our derivations and
verifying simulations. They obtained the  width $l(t)$ in their
numerical solution of the discrete SG in their Fig.(2) which we discussed in
detail in Sec.III. Furthermore, their Eq.(3) for $l(t)$ is incorrect because
it contains none of the many terms proportional to $\chi(t)$ that  appear in
the exact equation of motion for $\Gamma(t)$ in Eq.(8) which are necessary
for the existence of the quasimode. Consequently, their two equations of
motion for $P(t)$ and $l(t)$ which they state is the ``basis of  their
theoretical analysis'' of Ref.\cite{Quintero00} have absolutely no relevance
to our analytic derivation and confirming simulations.

One of the strangest aspects of Ref.\cite{Quintero00} is the complete lack of any mention 
or discussion of the continuum states $\chi$ of the SG equation in the presence of 
the SG soliton that are responsible for the existence of the SG quasimode.
The solution for $\chi$ derived in Eq.(11) of Ref.\cite{Boesch90} constitutes a dynamical 
dressing of the Sine-Gordon soliton due to the oscillation of $\Gamma(t)$.

Several times in Ref.\cite{Quintero00} the authors compare the SG quasimode
with the  $\phi^4$ equation internal mode of the which is an exact eigenmode
of the linearized $\phi^4$ equation  whose eigenfrequency is in the phonon
gap. They state that since they can  observe the $\phi^4$ eigenmode in
absorption but cannot observe the SG quasimode in  absorption, this proves
the SG quasimode doesn't exist. However, a quasimode  is not an eigenmode. A
quasimode can be observed in emission but cannot be  observed in absorption
unless it is first created and then observed within  its finite lifetime. In
Ref.\cite{Boesch90} we proved that the SG quasimode cannot be  excited by the
ac field used in Ref.\cite{Quintero00}.

Finally, in spite of the fact that all our derivations and verifying
simulations were done for the continuum SG, the simulations and analysis by
the authors of Ref.\cite{Quintero00} were done for an appreciably discrete SG
equation.  In particular they report finding a discrete mode in the phonon
gap that was  found by Kerekidis and Jones Ref.\cite{Kevrekidis00}. It is
important to observe that none  of the discrete simulations are in any way
relevant to the exact analytic  calculations and simulations of our continuum
SG equation.

\section{CONCLUSION}
In Ref.\cite{Quintero00} the authors did not make a simple reference to or
comment upon the exact analytic solutions and their verifying simulations of
the SG quasimode in Ref.\cite{Boesch90}. Furthermore, they never mention the
continuum SG states $\chi$ that interact with the SG soliton to form the SG
quasimode. The authors of Ref.\cite{Quintero00} performed two numerical
investigations of the discrete  SG equation to attempt to prove the SG
quasimode doesn't exist. The first simulation, an absorption measurement of
an ac field, which we proved could not possibly observe the SG quasimode in
principle. The second simulation was for an initially deformed discrete SG
soliton, which they followed in time. In the first 200 seconds they actually
observed the discrete SG soliton. However, they took a system too short for
their 2500 seconds observation time. Consequently, they observed a complicated
phonon interference pattern caused by the multiple interferences between
phonons emitted at different times which they incorrectly interpreted as the
nonexistence of the SG quasimode instead of the fact they observed the system
over twelve times too long for the length of their system.

Recently G. Kalbermann\cite{Kalbermann04} found new important analytic
nonperturbative solutions to the SG equation using the Inverse Scattering
Transform method. He states ``his solutions agree very closely with the
results of R. Boesch and C.R. Willis (our Ref.\cite{Boesch90}) in the
quasimode regime''as shown in his Fig.(4) of Ref.\cite{Kalbermann04}. Also,
Kalbermann in Ref.\cite{Kalbermann04} points out `` a probable source  of
error in the numerical calculations of Quintero et. al.\cite{Quintero00}.''
The numerical calculation he refers to in his Ref.\cite{Kalbermann04} is the
same numerical investigation which we labeled the Second Numerical
Investigation in Sec.IV  of the present paper. His explanation is essentially
identical to the explanation we give in Sec.IV of the present paper.

In their conclusion Quintero et.al.\cite{Quintero00} provided an explanation
of our results. They state ``the resonance
observed by Boesch and Willis took place in fact with the lowest frequency
phonon in the presence of a moving kink and not with any internal
quasimode.'' Their argument is, since the kink is moving, there is a Doppler
shift i.e.
$$
\bar{\omega}_k = \frac{\omega_k - k\,u(0)}{(1-u^2(0))^{1/2}},
$$
where
$$
\omega_k = (1+k^2)^{1/2}.
$$ 
However, all our analytic calculations and verifying simulations are
for a stationary kink i.e., $\dot{X}(t) \equiv 0$ so the kink is not
moving and thus the Doppler shift is identically zero. Therefore their
explanation that our results are due to a Doppler shift cannot
possibly be correct.


\begin{thebibliography}{10}

\bibitem{Boesch90} 
R. Boesch and C. R. Willis, Phys. Rev. B. {\bf 42}, 2290 (1990).

\bibitem{Boesch88}
R. Boesch and P. Stancioff and C. R. Willis, Phys. Rev. B. {\bf 38}, 6713
(1988). 

\bibitem{Quintero00}
N. R. Quintero and A. Sanchez and F. G. Mertens, Phys. Rev. E. {\bf 62}, R60
(2000).

\bibitem{Kevrekidis00}
P. G. Kevrekidis and C. K. R. T. Jones, Phys. Rev. E. {\bf 61}, 3114 (2000).

\bibitem{Kalbermann04}
G. Kalbermann, arXiv:cond-mat/0408198 (2004).

\end{thebibliography}
\end{document}